\documentstyle[aps]{revtex}
\draft

\begin{document}

\tighten

\twocolumn[\hsize\textwidth\columnwidth\hsize\csname@twocolumnfalse\endcsname

\title{Ordered valence bond states in symmetric two-dimensional spin-orbital
systems}
\author{Guang-Ming Zhang$^{1,2}$ and Shun-Qing Shen$^{1}$}
\address{$^{1}$Department of Physics, the University of Hong Kong, Pokfulam Road,
Hong Kong, China.\\
$^{2}$Center for Advanced Study, Tsinghua University, Beijing, 100084, China}
\date{\today}
\maketitle

\begin{abstract}
We consider a superexchange Hamiltonian, 
$H=-\sum_{<i,j>}(2{\bf S}_i\cdot {\bf S}_j-\frac 12)(2{\bf T}_i\cdot {\bf T}
_j-\frac 12)$,  which describes systems with
orbital degeneracy and strong electron-phonon coupling in the limit of large
on-site repulsion. In an SU(4) Schwinger boson representation, a reduced
spin-orbital interaction is derived {\it exactly}, and a mean field theory
has been developed by introducing a symmetric valence bond pairing order
parameter. In one dimension, a spin-orbital liquid state with a finite gap
is obtained. On a two-dimensional square lattice a novel type of spin-orbital
ferromagnetically ordered state appears, while spin and orbital are
antiferromagnetic. Moreover, an important relation has been found, relating
the spin and orbital correlation functions to the combined spin-orbital ones.
\end{abstract}


]

There have been much interests in the properties of Mott insulators with
orbital degeneracy \cite{wbao,mks,khaliullin,saitoh}, where the electron
configuration has an orbital degeneracy in addition to the spin degeneracy.
Due to the interplay between the spin and orbital degrees of freedom, a rich
variety of spin and orbital ordering effects have been displayed \cite
{kugel,feiner,aa95,pati,brink}, and a new combined spin-orbital degrees of
freedom may introduce some new physics to the transitional metal oxides \cite
{yqli,mila,joshi,azaria,ysu,affleck}. Recently, in order to describe the
low-energy physics of an insulating crystal with one-electron per site with
double orbital degeneracy in the limit of the large on-site Hubbard
repulsion, a superexchange Hamiltonian was proposed \cite{santoro1,santoro2} 
\begin{equation}
H=-\sum_{<i,j>}(2{\bf S}_i\cdot {\bf S}_j-\frac 12)(2{\bf T}_i\cdot {\bf T}%
_j-\frac 12),  \label{model}
\end{equation}
where two isotropic Heisenberg antiferromagnets (AFM) coupled by a quartic
term on equivalent bonds, and ${\bf S}_i$ and ${\bf T}_i$ denote the
spin-1/2 and orbital-1/2 operators at a lattice site $i$, respectively. The
physical condition to derive the above effective model Hamiltonian is that
among the possible two-particle states obtained upon virtual hopping, the
inter-orbital singlet is the lowest energy state due to the dynamical
Jahn-Teller effect. The condition might be realized in the large body of new
molecular compounds based on $C_{60}$ \cite{paul} or layered fullerides and
some two-dimensional copolymers, such as vinylidene
fluoridetrifluoroethylene \cite{choi}. Based on numerical calculations,
Santoro {\it et al. }argued that the ground state of the model shows a
spin-Peierls-like dimerization in one dimension (1D) \cite{santoro1}, while
on a two dimensional (2D) square lattice no evidence of order of any kinds
had been found that the ground state is concluded to be a spin liquid of
resonating valence bonds (VB) \cite{santoro2}. In order to put these results
on a firm ground, it is thus of great interest to develop new approaches to
this model.

In this Letter, we use an SU(4) Schwinger boson representation \cite{aa88}
to denote\ both the spin-1/2 and orbital-1/2 operators at the same time, and
the model Hamiltonian is simplified to a reduced interaction, describing an
symmetric pairing attraction among the nearest neighbor hard-core bosons. By
introducing an symmetric VB order parameter, a mean field theory is
developed. In 1D, a spin-orbital liquid state with a finite gap in the
excitation spectra is obtained, corresponding to a quantum {\it disordered}
VB state. On a 2D square lattice, due to the presence of the Bose-Einstein
condensations we found a novel spin-orbital ferromagnetically (FM) ordered
state, corresponding to a short-ranged VB crystal state, while both the spin
and orbital degrees of freedom form an AFM ordering.

The model Hamiltonian can be rewritten as follows 
\begin{eqnarray}
H &=&2\sum_{<i,j>}({\bf S}_i\cdot {\bf S}_j+{\bf T}_i\cdot {\bf T}_j) 
\nonumber \\
&&-\sum_{<i,j>}(2{\bf S}_i\cdot {\bf S}_j+\frac 12)(2{\bf T}_i\cdot {\bf T}%
_j+\frac 12),
\end{eqnarray}
where the direct quadratic couplings among spins and orbitals are
antiferromagnetic, respectively, while the quartic coupling between spins
and orbitals is ferromagnetic. First of all, the model Hamiltonian has an $%
SU(2)\otimes SU(2)$ symmetry, representing rotational invariance in both
spin and orbital space, and also interchange symmetry between the spins and
orbitals. Moreover, one also notices that the total spin, orbital, and
combined {\it staggered} spin-orbital operators 
\[
S^\alpha =\sum_jS_j^\alpha ,\text{ }T^\alpha =\sum_jT_j^\alpha ,\text{ }%
L^{\alpha \beta }=\sum_je^{i{\bf Q\cdot R}_j}2S_j^\alpha T_j^\beta , 
\]
where ${\bf Q}$ is the AF reciprocal vector, generate an SU(4) Lie algebra,
and also commutes with the model Hamiltonian. Therefore, Eq.(1) possesses
the SU(4) symmetry \cite{santoro1}. For a state with the SU(4) symmetry,
implying that the state is invariant under the rotation in the SU(4) space,
we have an identity for the static correlation functions

\begin{equation}
\left\langle S_{i}^{\alpha }S_{j}^{\alpha }\right\rangle =\left\langle
T_{i}^{\beta }T_{j}^{\beta }\right\rangle =e^{i{\bf Q\cdot }\left( {\bf R}%
_{i}-{\bf R}_{j}\right) }\left\langle 4S_{i}^{\alpha }T_{i}^{\beta
}S_{j}^{\alpha }T_{j}^{\beta }\right\rangle ,  \label{identity}
\end{equation}
where $\left\langle \cdots \right\rangle $ represents the expectation value
on the SU(4) symmetric state. The underlying physics for the identities is
very clear: if the spin and orbital degrees of freedom are
antiferromagnetically correlated, the combined spin-orbital correlation
functions will be ferromagnetic. This property originates from the
additional phase factor in the generators $L^{\alpha \beta }$. It should be
mentioned that another spin-orbital model Hamiltonian $H^{\prime
}=\sum_{<i,j>}(2{\bf S}_{i}\cdot {\bf S}_{j}+\frac{1}{2})(2{\bf T}_{i}\cdot 
{\bf T}_{j}+\frac{1}{2})$, the natural generalization of the SU(2)
Heisenberg spin model and {\it exactly} soluble in 1D, has also been used to
understand some orbital properties of transition metal oxides \cite
{yqli,mila,joshi,azaria,ysu,affleck}. However, although these two models
have the same symmetry, the fifteen generators of the SU(4) symmetry group
are not the same, and the identities of the correlation functions induced
from this symmetry are also different. In this sense, the physics involved
in these two models are independent, leading to different spin-orbital
properties of transitional metal oxides.

As we know, the Schwinger boson mean filed approach has been a successful
theory in describing the low-energy excitations of the conventional FM and
AFM Heisenberg spin superexchange models \cite{aa88,hirsch89}. In
particular, starting from short-ranged VB order parameters, it can produce
either quantum disordered or quantum long-range ordered magnetic states. In
the present model when two sets of SU(2) Schwinger bosons are introduced to
denote the spin-1/2 and orbital-1/2 operators separately, we have 
\begin{eqnarray*}
2{\bf S}_i\cdot {\bf S}_j+\frac 12 &=&\sum_{\alpha ,\beta }a_{i,\alpha
}^{\dagger }a_{i,\beta }a_{j\beta }^{\dagger }a_{j,\alpha } \\
2{\bf T}_i\cdot {\bf T}_j+\frac 12 &=&\sum_{m,n}d_{i,m}^{\dagger
}d_{i,n}d_{j,n}^{\dagger }d_{j,m}
\end{eqnarray*}
with local constraints $\sum_\alpha a_{i,\alpha }^{\dagger }a_{i,\alpha }=1$
and $\sum_md_{i,m}^{\dagger }d_{i,m}=1,$ where the indices $\alpha $, $\beta 
$ and $m$, $n$ are taken values 1 and 2 corresponding to eigenstates of $%
S^z=\pm 1/2$ and $T^z=\pm 1/2$, respectively. Both $a_{i,\alpha }$ and $%
d_{i,m}$ operators satisfy the boson commutation relations. To treat the
quartic spin-orbital interaction term on an equal footing as the same as the
spin-spin and orbital-orbital superexchange interactions, the fact has been
used that in the model Hamiltonian the Hilbert space on each lattice site
consists of four basic states in terms of $\left| S^z;T^z\right\rangle $:

\begin{eqnarray*}
\left| +\frac{1}{2};+\frac{1}{2}\right\rangle &=&\left| 1\right\rangle ,%
\text{ }\left| -\frac{1}{2};+\frac{1}{2}\right\rangle =\left| 2\right\rangle
, \\
\text{ \ }\left| +\frac{1}{2};-\frac{1}{2}\right\rangle &=&\left|
3\right\rangle ,\text{ }\left| -\frac{1}{2};-\frac{1}{2}\right\rangle
=\left| 4\right\rangle .
\end{eqnarray*}
These four states form a set of local basis to represent the SU(4) symmetry
group. The conventional SU(4) generators $S_{\mu }^{\nu }(i)$ act on a basic
state $|\eta \rangle _{i}$ according to the equation $S_{\mu }^{\nu
}(i)|\eta \rangle _{i}=\delta _{\nu ,\eta }|\mu \rangle _{i}$ with a local
constraint $\sum_{\mu }S_{\mu }^{\mu }(i)=1$, where the indices $\mu $ and $%
\nu $ are taken values from 1 to 4 corresponding to the four eigenstates of $%
\left| \pm \frac{1}{2};\pm \frac{1}{2}\right\rangle $. The SU(4) Lie algebra
is defined by the commutation relation 
\[
\left[ S_{m}^{n}(i),S_{k}^{l}(i)\right] =\delta _{n,k}S_{m}^{l}(i)-\delta
_{m,l}S_{k}^{n}(i). 
\]
In terms of four-component hard-core bosons, we have $S_{\mu }^{\nu
}(i)=b_{i,\mu }^{\dagger }b_{i,\nu }$ with the local constraint $\sum_{\mu
}b_{i,\mu }^{\dagger }b_{i,\mu }=1$. Therefore, the quartic spin-orbital
exchange interaction can be written as 
\begin{equation}
(2{\bf S}_{i}\cdot {\bf S}_{j}+\frac{1}{2})(2{\bf T}_{i}\cdot {\bf T}_{j}+%
\frac{1}{2})=\sum_{\mu ,\nu }b_{i,\mu }^{\dagger }b_{i,\nu }b_{j,\nu
}^{\dagger }b_{j,\mu },
\end{equation}
By a projection procedure, both the spin-spin and orbital-orbital quadratic
superexchange interactions can also be expressed in terms of the
four-component hard-core bosons. Fortunately, most of the resulting terms
are found to be {\it exactly} canceled with the quartic spin-orbital
exchange interaction, and the resulting Hamiltonian is reduced to a simple
and compact form

\begin{eqnarray}
H &=&-\sum_{<i,j>}\left[ (b_{i,1}^{\dagger }b_{j,4}^{\dagger
}+b_{i,4}^{\dagger }b_{j,1}^{\dagger })-(b_{i,2}^{\dagger }b_{j,3}^{\dagger
}+b_{i,3}^{\dagger }b_{j,2}^{\dagger })\right]  \nonumber \\
&&\text{ \ \ }\times \left[
(b_{j,4}b_{i,1}+b_{j,1}b_{i,4})-(b_{j,3}b_{i,2}+b_{j,2}b_{i,3})\right] ,
\end{eqnarray}
Actually, this reduced model Hamiltonian has explicitly displayed that the
main physics of the symmetrically coupled spin-orbital interaction is to 
{\it induce a unique attractive pairing instability among the nearest
neighbor hard-core bosons}. We would also emphasize here that the quartic
spin-orbital exchange interaction plays the {\it same} role as that of the
spin-spin and orbital-orbital superexchange terms, as both quadratic and
quartic exchange interactions have similar forms and are treated on the
equal footing.

In the reduced spin-orbital model Hamiltonian, the local attractive
interaction shares closely resemblance with the effective model Hamiltonian
in the conventional BCS superconductivity theory. Such a feature allows us
to introduce a unique short-ranged VB pairing order parameter

\[
\Delta =-\langle
(b_{j,4}b_{i,1}+b_{j,1}b_{i,4})-(b_{j,3}b_{i,2}+b_{j,2}b_{i,3})\rangle , 
\]
which can assume to be real. A mean field theory is followed to develop
naturally, leading to the mean field Hamiltonian

\begin{eqnarray*}
&&H_{mf}=\lambda \sum_{i,\mu }b_{i,\mu }^{\dagger }b_{i\mu }-\lambda
N+Z\Delta ^2N \\
&&\hspace{0.5cm}+\Delta \sum_{<i,j>}\left[ (b_{i,1}^{\dagger
}b_{j,4}^{\dagger }+b_{i,4}^{\dagger }b_{j,1}^{\dagger })-(b_{i,2}^{\dagger
}b_{j,3}^{\dagger }+b_{i,3}^{\dagger }b_{j,2}^{\dagger })+h.c.\right] ,
\end{eqnarray*}
where the local constraint has been imposed on average through a Lagrangian
multiplier, $N$ is the total number of lattice sites, and $Z$ is the number
of the nearest neighbors. In the momentum space, $H_{mf}$ becomes

\begin{eqnarray}
H_{mf} &=&\lambda \sum_{{\bf k,}\mu }b_{{\bf k},\mu }^{\dagger }b_{{\bf k}%
,\mu }-\lambda N+Z\Delta ^2N  \nonumber \\
&&+2Z\Delta \sum_{{\bf k}}\gamma _{{\bf k}}\left[ (b_{{\bf k,}1}^{\dagger
}b_{-{\bf k,}4}^{\dagger }-b_{{\bf k,}2}^{\dagger }b_{-{\bf k,}3}^{\dagger
})+h.c.\right] ,
\end{eqnarray}
where $\gamma _{{\bf k}}=\frac 1Z\sum_{{\bf \delta }}e^{i{\bf k\cdot \delta }%
}$ and the vector ${\bf \delta }$ points to the nearest neighbor sites. When
a Nambu spinor is defined by $\Psi ^{\dagger }({\bf k})=(b_{{\bf k}%
,1}^{\dagger },b_{{\bf k},2}^{\dagger },b_{{\bf k},3}^{\dagger },b_{{\bf k}%
,4}^{\dagger };b_{-{\bf k},4},b_{-{\bf k},3},b_{-{\bf k},2},b_{-{\bf k},1})$%
, the mean field Hamiltonian is expressed in a compact matrix form 
\begin{equation}
H_{mf}=\frac 12\sum_{{\bf k}}\Psi ^{\dagger }({\bf k})H_{mf}({\bf k})\Psi (%
{\bf k})+Z\Delta ^2N-3\lambda N,
\end{equation}
with $H_{mf}({\bf k})=\lambda \Omega _0+2Z\Delta \gamma _{{\bf k}}\Omega _2$%
. The corresponding Lagrangian is thus given by 
\[
L_{mf}=\frac 12\sum_{{\bf k}}\Psi ^{\dagger }({\bf k},i\omega _n)\left[
i\omega _n\Omega _1-H_{mf}({\bf k})\right] \Psi ({\bf k},i\omega _n). 
\]
Here the generalized $8\times 8$ Dirac matrices have been defined by $\Omega
_1=\sigma _z\otimes \sigma _0\otimes \sigma _0$, $\Omega _2=\sigma _x\otimes
\sigma _z\otimes \sigma _z$, where $\sigma _x$, $\sigma _y$, $\sigma _z$,
and $\sigma _0$ are three Pauli $2\times 2$ matrices and unity matrix,
respectively. $\Omega _0$ is the $8\times 8$ unity matrix, $\Omega _1$and $%
\Omega _2$ obey the anticommutation relation $\left[ \Omega _1,\Omega
_2\right] _{+}=2\Omega _0$. From the Lagrangian, the bosonic Matsubara Green
function is given by

\begin{equation}
G({\bf k,}i{\bf \omega }_n{\bf )=}\frac{-i\omega _n\Omega _1-\lambda \Omega
_0+2Z\Delta \gamma _{{\bf k}}\Omega _2}{\omega _n^2+\left[ \lambda
^2-(2Z\Delta \gamma _{{\bf k}})^2\right] },
\end{equation}
which implies that the bosonic quasiparticle excitations form a continuum
band, and their dispersion relation is $\omega _{{\bf k}}{\bf =}\sqrt{%
\lambda ^2-(2Z\Delta \gamma _{{\bf k}})^2}$. Moreover, the free energy can
be evaluated as 
\begin{equation}
F_{mf}=\frac 1\beta \sum_{{\bf k},\mu }\ln \left[ 2\sinh (\frac{\beta \omega
_{{\bf k}}}2)\right] +Z\Delta ^2N-3\lambda N,
\end{equation}
and the saddle point equations are derived as follows 
\begin{eqnarray}
\frac 1N\sum_{{\bf k}}\frac{2\lambda }{\sqrt{\lambda ^2-(2Z\Delta \gamma _{%
{\bf k}})^2}}\left[ 2n_B(\omega _{{\bf k}})+1\right] &=&3,  \nonumber \\
\frac 1N\sum_{{\bf k}}\frac{(2Z\gamma _{{\bf k}})^2}{\sqrt{\lambda
^2-(2Z\Delta \gamma _{{\bf k}})^2}}\left[ 2n_B(\omega _{{\bf k}})+1\right]
&=&Z.
\end{eqnarray}
The first equation corresponds to the request of the local constraint on
average $\sum_\mu \langle b_{i,\mu }^{\dagger }b_{i\mu }\rangle =1$, the
second one is to determine the short-ranged VB order parameter
self-consistently.

To get the collective excitations of the model, the dynamical correlation
functions of the spin, orbital, and combined spin-orbital density operators
should be calculated. The corresponding density operators can be written in
terms of the Nambu spinor introduced above,

\begin{eqnarray*}
S_{i}^{z} &=&\frac{1}{4}\Psi ^{\dagger }({\bf r}_{i})\Omega _{3}\Psi ({\bf r}%
_{i})\text{,} \\
T_{i}^{z} &=&\frac{1}{4}\Psi ^{\dagger }({\bf r}_{i})\Omega _{4}\Psi ({\bf r}%
_{i})\text{, } \\
L_{i}^{zz} &=&\frac{1}{4}\Psi ^{\dagger }({\bf r}_{i})\Omega _{5}\Psi ({\bf r%
}_{i})\text{,}
\end{eqnarray*}
where $\Omega _{3}=\sigma _{z}\otimes \sigma _{0}\otimes \sigma _{z}$, $%
\Omega _{4}=\sigma _{z}\otimes \sigma _{z}\otimes \sigma _{0}$, and $\Omega
_{5}=\sigma _{0}\otimes \sigma _{z}\otimes \sigma _{z}$ are introduced.
Except for \ the anticommutation relations $\left[ \Omega _{1},\Omega
_{2}\right] _{+}=\left[ \Omega _{2},\Omega _{3}\right] _{+}=\left[ \Omega
_{2},\Omega _{4}\right] _{+}=2\Omega _{0}$, all other relations between the
Dirac matrices from $\Omega _{1}$ to $\Omega _{5}$ satisfy the commutation
relations.

In 1D system, we have $Z=2$ and $\gamma _k=\cos k.$ At $T=0$K, there are no
particle excitations $n_B(\omega _k)=0$. When the summations over momenta
are converted into the integrals within the first Brillouin zone, the
self-consistent equations are easily solved, leading to $\Delta \simeq
1.67476$ and $\lambda \simeq 7.31391$. The corresponding ground state energy
per site is given by $\varepsilon _g=-2\Delta ^2\simeq -5.60964$, and there
is a finite gap in the bosonic quasiparticle excitation spectrum, $\omega _0=%
\sqrt{\lambda ^2-16\Delta ^2}\simeq 2.93533$ at the momentum $k=0$ and $%
k=\pi $. These results have shown that the ground state of the model is a
quantum {\it disordered} state of resonating valence bonds, in qualitative
agreement with the previous numerical calculations \cite{santoro1}.
Moreover, the dynamical susceptibilities at zero temperature can be easily
evaluated as follows

\begin{eqnarray}
&&\chi (q,\omega +i\eta )=\int \frac{dk}{16}\left[ \frac{\lambda ^2\pm
16\Delta ^2\cos k\text{ }\cos (k+q)}{\omega _k\omega _{k+q}}-1\right] 
\nonumber \\
&&\hspace{0.5cm}\times \left[ \frac 1{\omega +\omega _k+\omega _{k+q}+i\eta }%
-\frac 1{\omega -\omega _k-\omega _{k+q}+i\eta }\right] ,
\end{eqnarray}
where $\eta $ is a positive infinitesimal, both the spin and orbital dynamic
susceptibilities are equal to each other and choose the lower ``$-$''sign,
and the combined spin-orbital dynamic susceptibility takes the upper ``$+$''
sign. Then, the dynamic density spectra are easily obtained 
\begin{eqnarray}
%TCIMACRO{\func{Im} }
%BeginExpansion
\mathop{\rm Im}%
%EndExpansion
\chi _S(q &\rightarrow &Q,\omega )=%
%TCIMACRO{\func{Im} }
%BeginExpansion
\mathop{\rm Im}%
%EndExpansion
\chi _T(q\rightarrow Q,\omega )=%
%TCIMACRO{\func{Im} }
%BeginExpansion
\mathop{\rm Im}%
%EndExpansion
\chi _L(q\rightarrow 0,\omega )  \nonumber \\
&=&\frac 1{8|\omega |}\sqrt{\frac{4\lambda ^2-\omega ^2}{\omega ^2-4\lambda
^2+64\Delta ^2}}
\end{eqnarray}
where $2\sqrt{\lambda ^2-16\Delta ^2}<|\omega |<2\lambda $. Clearly, there
is a finite gap $E_r=2\omega _0\simeq 5.87066$ in both the spin and orbital
collective excitations at $q=Q$, and in the combined spin-orbital collective
excitation at $q=0$. Therefore we conclude that the ground state is a
quantum disordered spin-orbital liquid state of short-ranged valence bonds
with a finite gap in their collective excitations.

On a 2D square lattice, $Z=4$ and $\gamma _{{\bf k}}=\frac 12(\cos k_x+\cos
k_y)$. At zero temperature, the conversion from the summations over momenta
to the integrals will be {\it invalid} as $(8\Delta )\rightarrow \lambda .$
Following the analogous treatments for the Bose-Einstein condensation \cite
{huang}, we separate the divergent terms at ${\bf k}^{*}{\bf =0}$ and ${\bf k%
}^{*}{\bf =Q}$ from the summations to yield 
\begin{eqnarray}
\int \frac{d^2k}{(2\pi )^2}\frac \lambda {\sqrt{\lambda ^2-(8\Delta \gamma _{%
{\bf k}})^2}}+\frac{4n_B(\omega _{{\bf k}^{*}})}{\sqrt{1-(8\Delta /\lambda
)^2}} &=&\frac 32,  \nonumber \\
\int \frac{d^2k}{(2\pi )^2}\frac{\lambda \gamma _{{\bf k}}{}^2}{\sqrt{%
\lambda ^2-(8\Delta \gamma _{{\bf k}})^2}}+\frac{4n_B(\omega _{{\bf k}^{*}})%
}{\sqrt{1-(8\Delta /\lambda )^2}} &=&\frac \lambda {16}.
\end{eqnarray}
When $(8\Delta )\rightarrow \lambda $, the boson condensation happens, and
the ratio $\rho =4n_B(\omega _{{\bf k}^{*}})/\sqrt{1-(8\Delta /\lambda )^2}$
is finite, which can be determined from the first equation by setting $%
\lambda =8\Delta $ inside the integral: $\rho \simeq 0.107$. Similarly, from
the second equation one can also obtain the short-ranged VB order parameter $%
\Delta \simeq 1.3159$ and thus $\lambda \simeq 10.5271$. Moreover, the
ground state energy per site is also evaluated to be $\varepsilon
_g=-4\Delta ^2\simeq -6.9263$. The bosonic quasiparticle excitation spectrum
now becomes linearly dependence of momentum near the minimal points ${\bf k}%
^{*}{\bf =0}$ and ${\bf k}^{*}{\bf =Q}$. When the boson condensation is
carefully treated, the dynamical structure factors for spin, orbital, and
combined spin-orbital collective excitations can also be calculated through
the fluctuation dissipation theorem. After some algebra, the final results
are given by

\begin{eqnarray}
S_{S}({\bf q} &\rightarrow &{\bf Q},\omega )=S_{T}({\bf q}\rightarrow {\bf Q}%
,\omega )=S_{L}({\bf q}\rightarrow 0,\omega )  \nonumber \\
&=&\frac{\pi \rho ^{2}}{2}\delta (\omega )\text{sign}\omega +\frac{1}{\pi
\lambda }\text{K}\left( \frac{|\omega |}{2\lambda }\right) \sqrt{\left( 
\frac{2\lambda }{\omega }\right) ^{2}-1},
\end{eqnarray}
where K$(x)$ is the complete elliptical function of the second kind. In the
range of $0<|\omega |<2\lambda $, there is a sharp resonance in the
collective excitations at zero frequency, corresponding to the formation of
a novel long-range ordered spin-orbital crystal state of resonating valence
bonds. In such a state, both the spin and orbital degrees of freedom form an
AFM long-range order separately, while the combined spin-orbital degrees of
freedom has an FM long-range ordering. The corresponding magnetizations can
be read off from the coefficient of the delta function $m=\rho /\sqrt{8\pi }%
\simeq 0.021$, which is much {\it smaller} than that of the SU(2)
antiferromagnetic Heisenberg spin model. Actually, this result is consistent
with the identity in Eq.(\ref{identity}), as the VB state in the mean field
theory has kept the SU(4) symmetry of the model. The present results are 
{\it in conflict with} the prediction of a spin liquid state with a finite
energy gap by the quantum Monte Carlo Green function calculations on a {\it %
small} square lattice for the same model \cite{santoro2}. However, the very
small magnetization clearly indicates that the quantum fluctuations in the
spin-orbital coupled model are much stronger than those in the pure spin
model. The numerical data shows that the energy gap (even if it exists) is
very close to zero. In this sense, both analytical and numerical results
seem to be consistent with each other.

Several remarks are in order. i) The reduced model Hamiltonian has displayed
a unique form of attractive pairing instability among the nearest neighbor
hard-core bosons. Introducing a short-ranged VB pairing order parameter does
not break the global SU(4) symmetry of the model. ii) The mean field theory
in 1D has given rise to the correct ground state of a quantum disordered
spin-orbital liquid state, which was consistent with numerical results \cite
{santoro1}. iii) On a 2D square lattice, the long-range spin-orbital
ferromagnetic ordered state is obtained due to the Bose-Einstein
condensation. We can also expect that such a long-ranged ordered state
should be robust to the perturbations added even away from the SU(4)
symmetric point slightly. iv) Compared with the numerical method, the SU(4)
Schwinger boson mean field approach for the symmetrically coupled
spin-orbital systems has provided the insight on the exact nature of the
ground state and its first excitations.

In conclusion, we derive a reduced effective spin-orbital Hamiltonian in a
symmetric spin-orbital systems by introducing the SU(4) Schwinger boson
representation. After introducing a symmetric pairing VB order parameter, a
mean field theory has been developed, leading to a quantum disordered liquid
state with a finite gap in the collective excitations in 1D, and on a 2D
square lattice a novel long-range FM ordered spin-orbital crystal state
where both the spin and orbital degrees of freedom form AFM. The new
ordering properties due to the interplay between the spin and orbital
degrees of freedom can be detected in the future experiments on transitional
metal oxides.

This work was supported by a RGC grant of Hong Kong. G. -M. Zhang thanks
Prof. Lu Yu for stimulating his interest in this subject, and acknowledges
the partial support from NSF-China (Grant No. 10074036) and the Special Fund
for Major State Basic Research Projects of China (G2000067107).

\end{document}